\documentclass{nature}

\usepackage[sort&compress]{natbib}
\usepackage{aas_macros}
\title{Explosive lithium production
in the classical nova V339 Del (Nova Delphini 2013)}

\author{Akito\ Tajitsu$^{1}$, Kozo\ Sadakane$^2$, Hiroyuki\ Naito$^{3,4}$,
Akira\ Arai$^{5,6}$ \& Wako\ Aoki$^{7}$}

\newcommand{\kms}{km\,s$^{-1}$}
\newcommand\arcsec{\mbox{$^{\prime\prime}$}}%
\def\lesssim{\mathrel{\hbox{\rlap{\hbox{\lower4pt\hbox{$\sim$}}}\hbox{$<$}}}}
\def\gtrsim{\mathrel{\hbox{\rlap{\hbox{\lower4pt\hbox{$\sim$}}}\hbox{$>$}}}}
\def\apj{$Astrophys.\ J.$}
\def\apjl{$Astrophys.\ J.$}
\def\aj{$Astron.\ J.$}
\def\aap{$Astron.\ Astrophys.$}
\def\mnras{$Mon.\ Not.\ R.\ Astron.\ Soc.$}
\def\pasj{$Publ.\ Astron.\ Soc.\ Jpn$}
\def\iaucirc{$IAU Circ.$}

\begin{document}

\maketitle

\begin{affiliations}
 \item Subaru Telescope, National Astronomical Observatory of Japan, 650 North A'ohoku Place, Hilo, HI 96720, USA
 \item Astronomical Institute, Osaka Kyoiku University, Asahigaoka,
Kashiwara, Osaka 582-8582, Japan
 \item Graduate School of Science, Nagoya University, Furo-cho,
Chikusa-ku, Nagoya 464-8602, Japan
 \item Nayoro Observatory, 157-1 Nisshin, Nayoro, Hokkaido 096-0066,
 Japan  
 \item Center for Astronomy, University of Hyogo, Sayo-cho, Hyogo
679-5313, Japan
 \item Koyama Astronomical Observatory, Kyoto Sangyo University, Motoyama,
 Kamigamo, Kita-ku, Kyoto 603-8555, Japan 
 \item National Astronomical Observatory of Japan, 2-21-1 Osawa, Mitaka, Tokyo 181-8588, Japan
\end{affiliations}

\begin{abstract}
The origin of lithium (Li) and its production process have long been an
 unsettled question in cosmology and astrophysics.
Candidates environments of Li production events or sites suggested by previous
 studies include big bang nucleosynthesis, interactions of energetic
 cosmic rays with interstellar matter, evolved low mass
 stars, novae, and supernova explosions.
Chemical evolution models and observed stellar Li abundances
 suggest that at least half of the present Li abundance may have
 been produced in red giants, asymptotic giant branch (AGB) stars, and
 novae\cite{1999A&A...352..117R,2001A&A...374..646R,2012A&A...542A..67P}.
However, no direct evidence for the supply of Li from
stellar objects 
to the Galactic medium has yet been found.
Here we report 
 on the detection of highly blue-shifted resonance lines of the singly ionized
 radioactive isotope of beryllium, $^{7}$Be, in the near ultraviolet 
 (UV) spectra of the classical nova V339 Del (Nova Delphini 2013).
Spectra were obtained 38 to 48 days after the explosion.
$^{7}$Be decays to form $^{7}$Li within a short time (half-life 53.22 days
 \cite{2003NuPhA.729....3A}).
The spectroscopic detection of this fragile isotope implies that it has
 been created during the nova 
 explosion via the reaction $^{3}\mbox{He}(\alpha,\gamma)^{7}\mbox{Be}$   
 (ref.\,\bibpunct{}{}{,}{n}{}{;}\cite{1971ApJ...164..111C}\bibpunct{}{}{,}{s}{}{;}),
 and supports the theoretical  prediction that a significant amount of
 $^{7}$Li could be produced in classical nova explosions.  
This finding opens a new way to explore $^{7}$Li
 production in classical novae 
 and provides a clue to the mystery
 of the Galactic evolution of lithium.
\end{abstract}

V339 Del (= Nova Delphini 2013)  is a classical nova that was discovered
as a bright 6.8 magnitude (unfiltered) source 
on 2013 August 14.584 Universal Time ({\sc ut}) \cite{2013AAN...489....1W}.
After 40 hours from the discovery, a maximum was reached on Aug 16.25
(MJD = 56,520.25) at
$V = 4.3$ (ref.\,\bibpunct{}{}{,}{n}{}{;}\cite{2013ATel.5304....1M}\bibpunct{}{}{,}{s}{}{;}).
Then, it began a normal decline.

High-resolution spectra ($R = 90,000$--$60,000$) of V339 Del were obtained
at four epochs after its outburst (+38, +47, +48, and +52 d).
These spectra contain a series of broad emission lines originating from
neutral hydrogen 
(H\,{\sc i}, Balmer series) and other permitted transitions of neutral
or singly ionized species (e.g., Fe\,{\sc ii}, He\,{\sc i},
Ca\,{\sc ii}).
These emission lines are usually seen in post-outburst spectra of
classical novae.
Most of these broad emission lines are accompanied by sharp and
blue-shifted multiple absorption lines at their blue edges.
The typical radial velocity ($v_{\rm rad}$) of these highly
blue-shifted absorption lines
is $\sim-1,000$ \kms.
Figure\,1-a and -b display the spectrum obtained at +47 d
in the vicinities of H\,{\sc $\eta$} and Ca\,{\sc ii}\,K lines, respectively.
The H\,{\sc i} line is accompanied by a broad emission with a FWHM of 
$\sim$ 1,300 \kms\ centered at $v_{\rm rad} \sim 0$  \kms.
The Ca\,{\sc ii}\,K (and H) has a weak but broad emission
and strong absorption components caused by the interstellar (IS)
absorptions.
In addition to these profiles around the rest positions of both lines,
two sharp absorption components are found at 
$v_{\rm rad} = -1,268\pm2$ and $-1,103\pm1$ \kms.
Among these, the latter is
apparently stronger than the former. 
Such absorption line systems have been found in post-outburst spectra of several
classical novae 
\cite{2008ApJ...685..451W,2010PASJ...62L...5S}.
The absorption line systems in V339 Del 
contain numerous transitions originating from
singly ionized iron-group species (Fe\,{\sc ii}, Ti\,{\sc ii}, Cr\,{\sc ii},
Mn\,{\sc ii}, and Ni\,{\sc ii}).
The depths of all blue-shifted absorption lines in V339 Del are only $\lesssim25$\%\ of the
continuum, 
while the bottoms of some strong lines (e.g., Balmer lines; see
Extended Data Fig.\,2) have flat features suggesting that the absorption
is saturated.
These observational results can be interpreted as the effect of absorbing
materials partially covering the background light source. 
There are no Na\,{\sc i} D doublet lines, which are often found to be the
strongest absorption features in novae within a few weeks after 
their outbursts
\cite{2008ApJ...685..451W,1960stat.conf..585M}.  
We interpret this as indicating that the ionization state of the ejected
gas has evolved into a higher stage of excitation before our observing
epochs (5--7 weeks). 
The observed spectral energy distribution of this nova indicates that
the shape of the continuous radiation had entered a very hot stage
(effective temperature $>$100,000 K) within 5 weeks after the
explosion\cite{2014A&A...569A.112S}. 
Other observed characteristics of this nova (e.g., light curves,
optical and UV emission lines) show that it is a typical Fe\,{\sc
ii} nova with a CO white dwarf (WD)
\cite{1992AJ....104..725W,1995CAS....28.....W}. 

Among these absorption line systems, we have noticed two remarkable pairs of 
absorption features near 312 nm. 
These correspond to the absorption
components originating from transitions at $\sim$313 nm. 
These pairs are marked as  A, B and C, D, respectively, in
 Figure\,1-c.
Adopting the wavelengths of the resonance doublet lines of singly
ionized $^{7}$Be at 313.0583 and 313.1228 nm \cite{NIST_ASD,2008PhRvL.100x3002Y}, 
we find that  features A and B coincide with the
$-1,103$ \kms\ components of  the 313.0583 and  313.1228 nm lines, 
respectively.
Similarly,  features C and D coincide with the
$-1,268$ \kms\ components of  these two lines.
Separations in wavelength between features A and B, and between features
C and D
are consistent with the separation between the doublet lines within the
measurement uncertainties.
Figure\,1-d illustrates these coincidences on the
velocity scale.  
Thanks to the high resolution of the spectrum ($\sim0.0052$ nm),
we can clearly distinguish them from the doublet of $^{9}$Be\,{\sc ii} 
at 313.0422 and 313.1067 nm \cite{NIST_ASD}.
After ruling out the possibilities of alternative identifications,
we conclude that these absorption features at 312 nm are caused by
$^{7}$Be, and not by $^{9}$Be which is the only stable
isotope of Be.
Original $^{9}$Be contained in the progenitor star would have been depleted 
during its evolution because this isotope is destroyed at temperatures 
$T > 3\times10^{6}$ K.
On the other hand, production of the unstable isotope $^{7}$Be
by the reaction $^{3}\mbox{He}(\alpha,\gamma)^{7}\mbox{Be}$  
in nova explosions has been theoretically predicted
\cite{1975A&A....42...55A,1978ApJ...222..600S,1991A&A...248...62D,1993A&A...279..173B,1996ApJ...465L..27H,1998ApJ...494..680J}.

The transition probability of the $^{7}$Be\,{\sc ii} 
line at 313.0583 nm ($\log{gf} = -0.178$)
is twice as large as that of the
$^{7}$Be\,{\sc ii} at 313.1228 nm ($\log{gf} = -0.479$) \cite{NIST_ASD}.
Due to saturation effects, 
the ratio of their equivalent widths is expected to be in the range between 
$2$ (no saturation) to $1$ (complete saturation).
The measured ratios are $1.1 \pm 0.3$ and $1.6 \pm 0.4$
for the components at $v_{\rm rad} = -1,268$ and $-1,103$ \kms,
respectively.
These are within the range expected for the doublet, although the values
contain some errors ($\lesssim\pm25$\%)
due mainly to the uncertainty in the continuum placement.
The weaker component at $v_{\rm rad} = -1,268$ \kms\ has a ratio closer
to complete saturation.
This can be interpreted as resulting from the fact that the absorbing
gas cloud moving with  
$v_{\rm rad} = -1,268$ \kms\ has a smaller covering factor, and 
at the same time, higher column density of $^{7}$Be ion than the gas
cloud moving with 
$v_{\rm rad} = -1,103$ \kms.

Figure\,2 displays the velocity plots of normalized spectra
for different species of the absorption line systems at four
observing epochs from +38 to +52 d. 
On +38 d, the $^{7}$Be\,{\sc ii} doublet has an absorption component at
$v_{\rm rad} = -1,386 \pm 3$ \kms and shows a complicated profile near $v_{\rm
rad} \sim-1,000$ \kms. 
From +47 to +48 d, the absorption components at $v_{\rm rad} = -1,268$ and
$-1,103$ \kms\ in +47 d shift by $-26 \pm 3$ and $-17 \pm 4$ \kms\
blueward, respectively.
These changes can be interpreted as due to the fact that we are
observing accelerating blobs of nova ejecta.
All of the blue-shifted absorption line systems had disappeared in the spectrum of + 52 d
except for the meta-stable He\,{\sc i} lines at 318.8 and 388.9 nm.
This fact indicates that the gas in the absorption line systems have evolved into a higher
ionization stage as discussed in the case of the nova V1280 Sco
\cite{2013PASJ...65...37N}.
These observations show -- 
(1) 
Several blue-shifted absorption lines with different velocities are
found from different species at each epoch;
(2)
Radial velocities of different transitions belonging to a velocity
component determined by Gaussian fittings agree within $|\Delta v_{\rm rad}| <$1--3 \kms;
(3)
Each component shifts blueward with time, 
indicating that the ejecta is being accelerated;
(4)
The strengths of the blue-shifted absorption lines weaken quickly
during the observing period.
The velocities and the strengths of the $^{7}$Be\,{\sc ii} doublet 
behave perfectly synchronized with those of other species.
This means that the gas producing the absorption line systems of V339
Del contains a considerable amount of $^{7}$Be ion which can produce
detectable absorption lines and 
strongly suggests that the gas  must have experienced an explosive
thermonuclear runaway on the surface of the WD.

Our spectroscopic detection of $^{7}$Be in a
classical nova immediately connects to the production of $^{7}$Li.
The production of $^{7}$Be via the nuclear reaction
$^{3}\mbox{He}(\alpha,\gamma)^{7}\mbox{Be}$ in novae 
has been studied theoretically.
However, no observational confirmation has been made up to the present
time. 
This is because the presence of $^{7}$Be is very transient and is only
observable in the near UV range where the atmospheric absorption
severely obstructs observations from ground-based telescopes.
In the case of V339 Del, the $^{7}$Be doublet can be identified only
within a very short period (from $\sim$6 to $\sim$7 weeks after the outburst).
In earlier epochs, it might be difficult to identify due to saturation effects.
For nearby bright novae, there have been several attempts to detect the
478 keV gamma-ray line produced by the decay of $^{7}$Be.
However, no definite detection has been reported because of the insufficient
sensitivity\cite{2001ApJ...563..950H,2008clno.book.....B}.
The $^{7}$Be absorption lines in the near UV spectrum of V339 Del are
found in highly blue-shifted ($\sim1,000$ \kms) flows, which have been
blown off by the outburst.
This means that it will soon decay to $^{7}$Li in
cooler interstellar or circumstellar matter on a time scale given by the 
half-life of $^{7}$Be (53.22 days).
The absence of the $^{7}$Li\,{\sc i} line at 670.8 nm in our spectra can be
interpreted as due to the fact that all of Li in 
the absorbing material of V339 Del 
has been ionized during the observing period as mentioned above.
This is in accordance with the fact that no Na\,{\sc i}\,D lines are found
in the absorption line system. 

The $^{7}$Be\,{\sc ii} doublet corresponds to the doublet of the
Ca\,{\sc ii} resonance lines (H and K lines) on the atomic energy level diagrams.
The Ca\,{\sc ii}\,K line at 393.366 nm has $\log{gf} = +0.135$ (ref.\,\bibpunct{}{}{,}{n}{}{;}\cite{NIST_ASD}\bibpunct{}{}{,}{s}{}{;}).
Supposing that most of the $^{7}$Be and Ca in the absorption line system
are singly 
ionized and the resonance lines of both ions are unsaturated,
the ratio of their equivalent widths directly reflects the number
density ratio between $^{7}$Be and Ca ions.
In the spectrum obtained at +47 d,
the ratios of $^{7}$Be\,{\sc ii} (313.1228 nm)/Ca\,{\sc ii}\,K, which
are less affected by saturation and/or contamination, are $\sim1.3 \pm 0.3$ and
$\sim0.7 \pm 0.2$ for the $-1,268$ and the $-1,103$ \kms\ components,
respectively.
This means that the column number density of $^{7}$Be is 5.3--2.9
times higher than that of Ca.
Using this $^{7}$Be/Ca number ratio, the mass fraction of $^{7}$Be
relative to the sum of all the constituent mass components, 
$X(^{7}{\rm Be})$, is presented as  
$(4.4\pm2.2) \times 7 / 40 \times X({\rm Ca})$.
If we adopt the solar $X({\rm Ca})$ ($= 10^{-4.19}$), for instance, 
the  $X(^{7}{\rm Be})$ in the absorbing gas system should amount to
$\sim10^{-4.3\pm0.3}$. 
The error estimation includes the uncertainty in the local continuum
placement ($\lesssim\pm$25\%) and the difference of derived equivalent
widths derived from individual velocity components ($\sim\pm$30\%).
Several additional factors, such as the abundance of Ca in the absorbing
gas,
and the difference in the ionization state between $^{7}$Be and Ca, are
difficult to estimate because the nova ejecta model is not yet established.
Taking such uncertainties into account, the error involved in the
above estimate of the $^{7}$Be abundance 
could be even larger by a factor of several.
However, in spite of the large uncertainty,
the abundance of $^{7}$Be is larger than, or at least as large
as, theoretical predictions for CO novae
[e.g., $X(^{7}{\rm Be}) \lesssim10^{-5.1}$
(ref.\,\bibpunct{}{}{,}{n}{}{;}\cite{1998ApJ...494..680J}\bibpunct{}{}{,}{s}{}{;})].
This indicates that classical novae could play an important role as 
contributors of $^{7}$Li in the Galaxy.

The observed $^{7}$Li evolutionary curve\cite{2012A&A...542A..67P} has a
plateau for young Galactic ages ($\lesssim2.5$ Gyr) followed by a steep rise.
To explain this requires a relatively
low-mass stellar component that evolves over a long lifetimes.
Candidates for this,
such as low-mass red giants or novae, have been proposed to be 
major sources of $^{7}$Li production
($> 50$\%\ of the solar system Li measured in meteorites) 
in the Galaxy
\cite{1999A&A...352..117R,2001A&A...374..646R,2012A&A...542A..67P}. 
The production of $^{7}$Li in low-mass stars has been theoretically studied
\cite{1999ApJ...510..217S,2000ApJ...535L.115D,1992ApJ...392L..71S,2001ApJ...559..909T,2010MNRAS.402L..72V},
and Li-enhanced red giants and AGB stars are indeed
identified\cite{2005A&A...439..227M}. 
The contribution to Li enrichment in the Galaxy by these
objects has, however, not been confirmed.
The reason for this is that the Li-rich phase in these
stars might be of quite limited duration and the contribution is
dependent upon the mass-loss rate of such objects.
Nova eruptions involve  a long delay time before working 
as stellar $^{7}$Li factories.
This is because $^{3}$He rich low mass
secondaries are necessary to produce $^{7}$Be efficiently via the 
$^{3}\mbox{He}(\alpha,\gamma)^{7}\mbox{Be}$ reaction
\cite{1991A&A...248...62D}. 
It is important to know whether this phenomenon is common
among classical novae to quantify their contribution to the rapid increase
of $^{7}$Li in the Galaxy.
Since V339 Del appears to be one of the ordinary Fe\,{\sc ii} type novae
which occupy $\sim$60 \%\ of all classical novae 
\cite{1992AJ....104..725W},
the $^7${Be} production found in this object might be 
occurring in many classical novae.
Our successful detection of $^{7}$Be in V339 Del indicates that
measurements of the $^{7}$Be lines in the near UV range for
post-outburst novae within the lifetime of this isotope is a powerful
way to estimate the contribution of novae to the chemical evolution of
lithium in the Galaxy. 




\begin{addendum}
 \item
 This work is based on data collected at Subaru Telescope,
 which is operated by the National Astronomical Observatory of Japan
 (NAOJ).
 We acknowledge with thanks the variable star observations from the AAVSO
 International Database contributed by observers worldwide and used in
 this research. 
 \item[Author Contribution] 
 A.T. planned and  carried out  Subaru HDS observations, reduced and 
 analyzed the data and  prepared the manuscript. 
 K.S., H.N., A.A., and  W.A.
 participated in the discussion and contributed in the process of  
 manuscript preparation significantly.
 \item[Competing Interests] The authors declare that they have no
competing financial interests.
 \item[Correspondence] Correspondence and requests for materials
should be addressed to A.T.~(email: tajitsu@naoj.org).
\end{addendum}

\newpage

\begin{figure}
 \includegraphics[width=8cm]{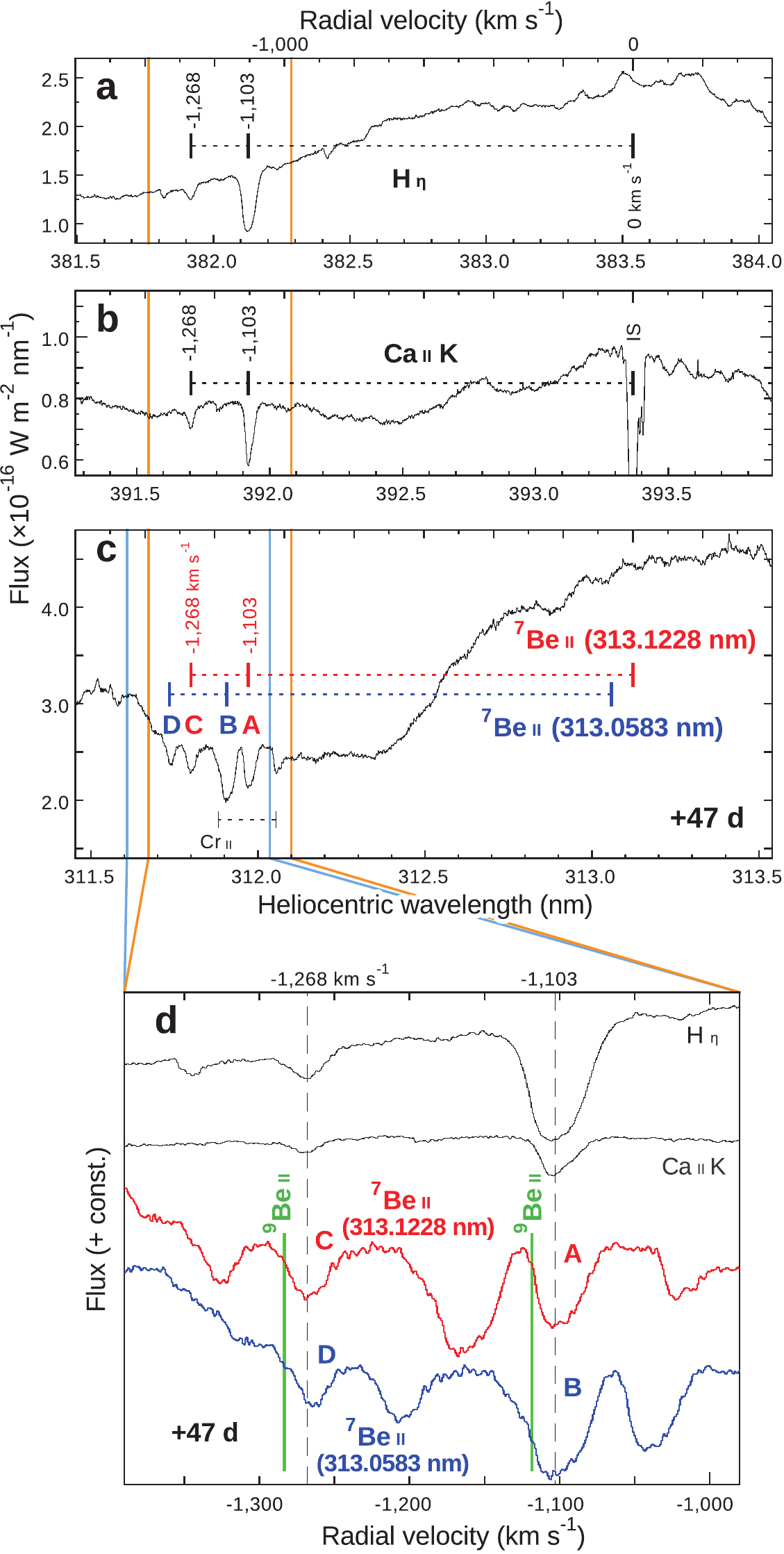}
 \caption{{\bf Blue-shifted absorption line systems in the spectrum
 of V339 Del obtained at +47 day.}
 Panels {\bf a}--{\bf c} display  
 the spectrum in the vicinity of H\,{\sc $\eta$} ({\bf a}), Ca\,{\sc ii}\,K ({\bf b}), 
 and the  $^{7}$Be\,{\sc ii} doublet ({\bf c}),
 on the velocity (upper horizontal) scale.
 Two blue-shifted absorption components and the zero velocity position for
 each line are indicated by ticks.
 {\bf c}, The velocity scale is adjusted to one of the $^{7}$Be\,{\sc
 ii} doublet (313.1228 nm, red).
 The positions of blue-shifted components of Cr\,{\sc ii}
 at 313.205 nm are displayed at the bottom.
 Panel {\bf d} shows the enlarged radial velocity profiles.
 The vertical dashed lines show two common absorption components.
 The expected positions of the $^{9}$Be\,{\sc ii} doublet are indicated by green lines.
}
\end{figure}

\begin{figure}
 \includegraphics[width=17cm]{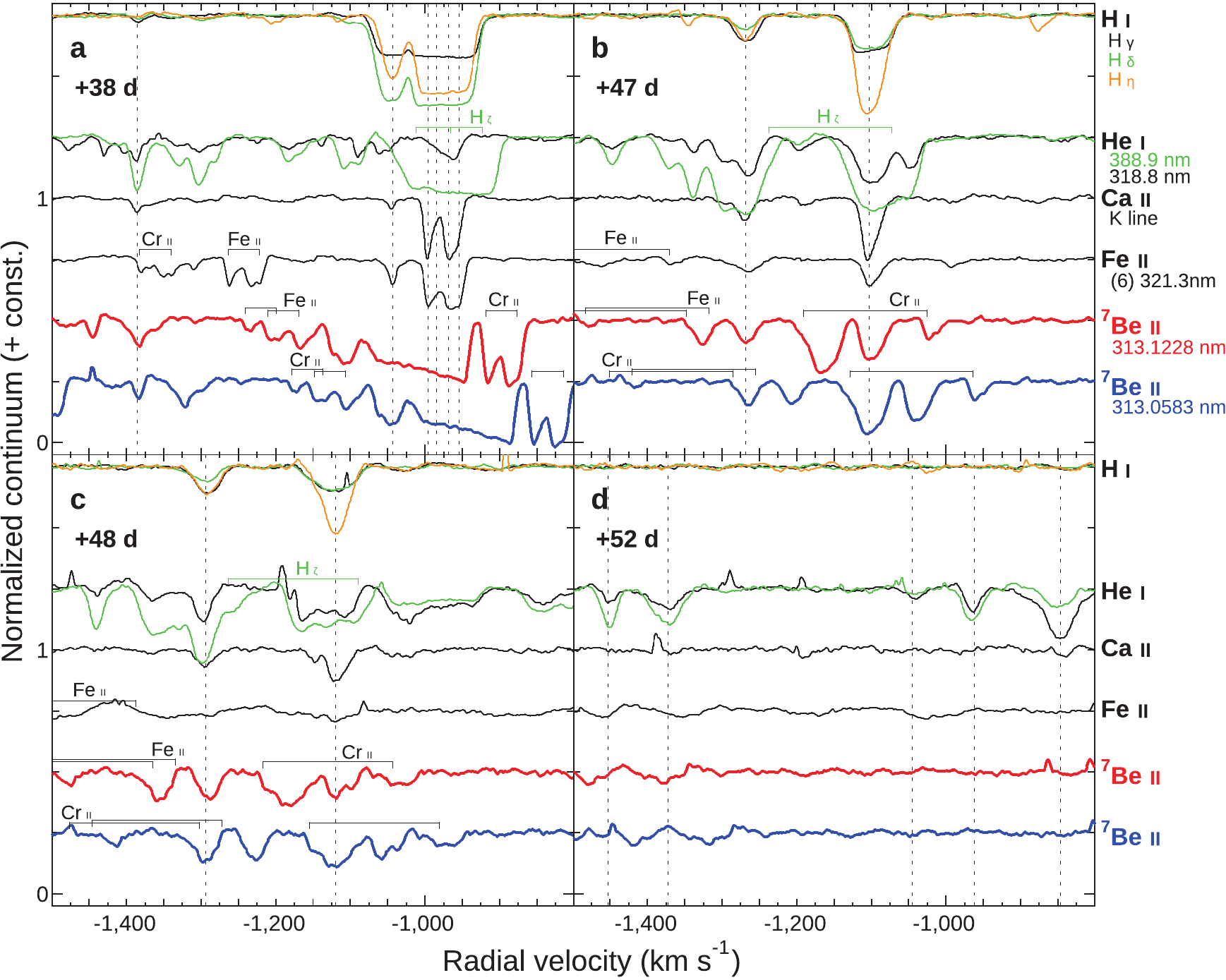}
\caption{
{\bf Time variations of the blue-shifted absorption line systems from +38 to +52 d.} 
 Absorption line systems originating from different species at four observing epochs 
 are plotted on the velocity scale.
 All lines are normalized to the local continuum.
 Blue-shifted absorption components observed at each epoch are indicated
 by vertical dashed lines. 
 The identified line or expected line contaminations are labeled above each
 lines.
 {\bf a}, On +38 d, the profile of the $^{7}$Be\,{\sc ii} doublet around 
 $v_{\rm rad} \sim-1,000$ \kms\ is complicated,
 and possibly interpreted as being saturated.
 {\bf d}, On +52 d, no blue-shifted absorption can be found
 except for the metastable He\,{\sc i} lines.
}
\end{figure}

\clearpage


\begin{methods}

\subsection{Discovery}
V339 Del (= Nova Delphini 2013) is a classical nova that was discovered
as a bright 6.8 magnitude (unfiltered) source by Koichi Itagaki
on 2013 August 14.584 {\sc ut} and announced in American Association of
Variable Star Observers (AAVSO) Alert Notice 
\cite{2013AAN...489....1W}.
Its progenitor is estimated 
 to be a blue star USNO B-1 1107-0509795  
with $B \sim17.20, R_{\rm C} \sim17.45$ on the first Palomar Sky
Survey Plates (exposed on 1951 July 7), 
and with $B \sim17.39, R_{\rm C} \sim17.74$ on the second Palomar
Sky Survey plates (exposed on 1990 July 18 and September 15,
respectively) \cite{2013IBVS.6087....1M}.
No significant changes were found in its
photometric behavior for at least a few years prior to the outburst 
\cite{2014A&A...563A.129D}.
On an unfiltered pre-discovery image obtained on 2013 August 13.998 
{\sc ut}, 
the object was still at 17.1 mag \cite{2013IAUC.9258....2D}.
This means that the object was still in quiescence until at least 14 hours
before its discovery, and that it showed a very fast rise to the maximum.
After 40 hours from the discovery, maximum was reached on Aug 16.25
(MJD = 56,520.25) at
$V = 4.3$
(ref.\,\bibpunct{}{}{,}{n}{}{;}\cite{2013ATel.5304....1M}\bibpunct{}{}{,}{s}{}{;}).
Then, it began a normal decline.
The nova had been detected as a transient high energy gamma ray ($>$
100 MeV) source within $\sim$10 days after the outburst
\cite{2014Sci...345..554H}.
Angular sizes of the expanding shell around the nova had been monitored
until $\sim$+40 d using near-infrared interferometric 
observations\cite{2014Natur.Schaefer}. 
Then, combining the expansion velocity obtained in the optical region,
the distance to the nova had been derived as 4.54 $\pm$ 0.59
kiloparsecs from the Sun.

\subsection{Observations and data reduction}
Post-outburst spectra of V339 Del were obtained using the
High Dispersion Spectrograph (HDS) \cite{2002PASJ...54..855N}
 of the 8.2 m Subaru Telescope at four epochs from 2013 September 23  
to October 7 (+38, +47, +48, and +52 d after the maximum).
According to the AAVSO light curves (see in Extended Data Fig.\,1),
our first observation was just before the start of the rapid decline in
optical magnitudes by dust formation \cite{2013ATel.5431....1S}.
The following three were obtained during the continuous decline.
We obtained spectra under 3 configurations of the spectrograph,
which cover the wavelength regions from 303 to 463 nm,
from 411 to  686 nm, and from 667 to 936 nm.
Spectral resolving power was set to $R \simeq90,000$ or 60,000 with
0\arcsec.4 (0.2 mm) or 0\arcsec.6 (0.3 mm) slit widths,  
respectively.
The exact times and wavelength ranges of obtained spectra are summarized
in Extended Data Table 1.
Data reduction was carried out using the IRAF software in a standard manner.
The non-linearity in the detectors are corrected by 
the method described in ref.\,\bibpunct{}{}{,}{n}{}{;}\cite{2010PNAOJ...13..1}\bibpunct{}{}{,}{s}{}{;}.
The wavelength calibration has been performed using a Th-Ar comparison
spectrum and the typical residual in wavelength calibration is
$\lesssim10^{-4}$ nm ($\sim$0.1 \kms) for each spectrograph configuration.
The typical systemic variance of the spectrograph is $\lesssim10^{-4}$ nm
per an hour.
We also examined the accuracy of radial velocity determination in our
measurement using the identified iron-group transitions in 315--351 nm,
For the spectrum obtained at +38d, the velocity of the strongest
component in the absorption line system was  $-996.1 \pm 0.7$ \kms\ 
determined by Gaussian fittings. 
In total,  we concluded that the residual in our velocity scale determination
was $\sim\pm$1 \kms.
Spectrophotometric calibration was performed using the spectrum of
BD+28$^{\circ}$ 4211 (ref.\,\bibpunct{}{}{,}{n}{}{;}\cite{1988ApJ...328..315M}\bibpunct{}{}{,}{s}{}{;})
obtained nearly at the same altitude of the nova
on the same nights.
All spectra were converted to the heliocentric scale.
Correction for interstellar extinction has not been applied.
The average signal-to-noise ratio in the spectra obtained at four epochs
is $\sim$140 at $\sim$312 nm, where we found the $^{7}$Be lines.

\subsection{Highly blue-shifted absorption line system}
The spectra of V339 Del exhibit a series of broad Fe\,{\sc ii} emission
lines, which indicate that the object is a typical Fe\,{\sc ii} type nova
\cite{1992AJ....104..725W}.
Since no strong emission originating from Ne is found in the
spectrum even in +52 d, 
the WD in the system is supposed to be a CO-WD \cite{1995CAS....28.....W}.

Extended Data Fig.\,2-a displays the radial velocity
profiles of three Fe\,{\sc ii} belonging to multiplet number
\cite{1959mtai.book.....M} (42) 
in the spectrum of +38 d.
The absorption line system on +38 d clearly consists of five components at  
$v_{\rm rad} = -954, -968, -985, -996, $ and $-1,043$ \kms\ as indicated by
the dashed lines in Extended Data Fig.\,2-b.
Similarly, blue-shifted absorption components are found in Balmer lines
(Extended Data Fig.\,2-c) 
and also in other permitted lines (Ca\,{\sc ii}\,H and K, He\,{\sc i} at 587.6 nm).
In the near UV range, numerous absorption lines in the complex
continuum are identified as the transitions of singly ionized iron-group
species.
Most of them belong to the absorption line systems found in the visual region
(Extended Data Fig.\,3). 
We applied a Doppler correction using the the radial velocity of the
strongest blue-shifted absorption line in the system to identify sources
of transitions (see in Extended Data Fig.\,3-b).
We use the velocities for Doppler corrections as 
$v_{+38} = -996$, $v_{+47} = -1,103$, 
and $v_{+48} = -1,120$ \kms\ for +38, +47, and +48 d, respectively
(in Extended Data Fig.\,3-b and Fig.\,4-a, b, and c).
All of the identified transitions originate from levels of
low excitation potentials ($\lesssim$4 eV).
The residual intensity at the bottom of these lines exceeds 75 \% of the
continuum, while the bottoms of some strong Fe\,{\sc ii} and Balmer lines show flat
structures.
These observations suggest that the saturated absorption lines are
created by clouds of absorbing gas, which cover the continuum emitting
region by only about 25 \%. 

Very similar short-lived blue-shifted metallic absorption systems
 (Transient Heavy Element Absorption; THEA)
 have been reported in post-outburst spectra of several classical novae\cite{2008ApJ...685..451W}.
Especially, great majority of novae show strong blue-shifted (400--1,000
 \kms) multiple absorption components in the 
Na\,{\sc I}\,D doublet in days following
their outbursts.
In the very slow nova V1280 Sco, 
multiple high-velocity (700--900 \kms) absorption components have been
found in the Na\,{\sc I} D doublet even 800 days after its maximum 
\cite{2010PASJ...62L...5S}.
Although no absorption components of the
Na\,{\sc I} D are found at all epochs of our observations of V339 Del,
other characteristics of the absorption line systems found in V339 Del are quite similar to
those of the THEAs in other novae.
They are 
(a) highly blue-shifted ($\sim1,000$ \kms),
(b) divided into several velocity components,
(c) time variable in their shapes and velocities,
and (d) short-lived (2--8 weeks).

\subsection{Contamination to the $^{7}$Be\,{\sc ii} doublet}
We carefully inspected possible contaminations of absorption lines
originating from other species to the $^{7}$Be\,{\sc ii} lines
consulting the atomic line database\cite{1995KurCD..23.....K}.
Extended Data Fig.\,4 displays the spectra in the
vicinity of the $^{7}$Be\,{\sc ii} doublet obtained at four epochs of
our observations. 

On the spectrum obtained at +47 d, 
there are no candidates to contaminate 
the $-1,103$ or the $-1,268$ \kms\ components of $^{7}$Be\,{\sc ii} at 
313.1228 nm, which we use in our $^{7}$Be abundance estimation 
(see in Extended Data Fig.\,4-b).

At this epoch, the other line of the $^{7}$Be doublet at 313.0583 nm
may be contaminated by some lines originating from iron-group species.
We estimate that the $-1,268$ \kms\ component of Cr\,{\sc ii}\,(5) at
313.205 nm ($\log{gf} = +0.079$) may contaminate to the $-1,103$ \kms\
component of this $^{7}$Be\,{\sc ii} line.
The influence of this contamination can be evaluated adopting the line
 strength ratio between the pair of velocity components
 of Cr\,{\sc ii}\,(5) at 312.497 nm ($\log{gf} = +0.018$)
 to that of Cr\,{\sc ii}\,(5) at 313.205 nm.  
It is quite small compared with the strength of the $^{7}$Be\,{\sc ii}
line ($\lesssim5$ \%).
Concerning the $-1,268$ \kms\ component of this $^{7}$Be\,{\sc ii} line,
we conclude that the weak lines of Fe\,{\sc ii}\,(96) at 312.901
nm ($\log{gf} = -2.70$) 
and Cr\,{\sc ii}\,(5)\ at 312.869 nm ($\log{gf} = -0.32$) do not
contaminate severely. 
This is because similar weak lines of Fe\,{\sc ii}\,(82) at 313.536 nm
($\log{gf} = -1.13$) and Cr\,{\sc ii}\,(5) at 313.668 nm  
($\log{gf} = -0.25$) had completely disappeared until + 47 d.
We can neglect the contamination from the V\,{\sc ii}\,(1) line at
313.0257 nm ($\log{gf} = -0.29$),
because the other V\,{\sc ii}\,(1) line at 312.621nm ($\log{gf} =
-0.27$) is  not detected on our spectra. 

\subsection{$^{7}$Be abundance estimation}
We empirically estimate the abundance of $^{7}$Be in the absorbing gas by
comparing the equivalent widths of the $^{7}$Be\,{\sc ii} line with 
those of the Ca\,{\sc ii}\,K line that are the similar transitions on the
atomic energy level diagrams.
We assume that the covering factor of the absorbing gas cloud to the
background illuminating source has no wavelength dependence.
This method could be a simple and robust approach to estimate the abundance 
ratio independent of ejecta models for nova explosions. 
The estimate, however, includes some uncertainties.
One is the difference in the ionization potentials between Be
(the first and second ionization potentials; $I_{1} = 9.32$,
$I_{2} = 18.21$ eV) and Ca
($I_{1} = 6.11$, $I_{2} = 11.87$ eV) \cite{NIST_ASD}
 that
could result in a difference of ionization states between Be and Ca.
However, all of iron-peak elements (Ti to Fe) found in the absorption line systems, 
which have intermediate ionization potentials ($I_{1} =$ 6.75--7.90,
$I_{2} =$ 13.58--18.12 eV) between those of Be and Ca, 
are observed  only in singly ionized states, 
suggesting that dominant fractions of Be and Ca are in the singly
ionized states, too.
In the obtained spectra, we could not find any resonance lines 
of Sr\,{\sc ii} or Ba\,{\sc ii}, which correspond to those of
Be\,{\sc ii} and Ca\,{\sc ii}. 
The Sr/Ca and the Ba/Ca number ratios would be quite small, as seen in
the solar abundance ($\ll 10^{-3}$).
Another uncertainty is the $X({\rm Ca})$
in the absorption line system.
Our assumption that the absorbing gas has the solar $X({\rm Ca})$
would not
be far from the reality, because the theoretical analysis predicts no
overabundance of elements with the mass number $>$30 in ejecta
of CO novae\cite{1998ApJ...494..680J}.
We remark that our $^{7}$Be abundance estimation is carried out using 
the data obtained at +47 d, which is close to the half-life of $^{7}$Be 
(53.22 days). 
Therefore, the abundance of the freshly produced $^{7}$Li 
in this nova explosion could be $\sim$2 times higher than the
$X(^{7}\mbox{Be})$ on +47 d.

\end{methods}

\newpage

\noindent
Extended Data Table 1:
{\bf Journal of HDS observations of V339 Del}\\

\noindent
Extended Data Fig. 1:
 {\bf Optical light curves of V339 Del.}
 $V$ (green) and $R$ (red) magnitudes are taken from the AAVSO database.
 The epochs of our HDS observations are indicated by arrows.
\\

\noindent
Extended Data Fig. 2:
 {\bf Optical spectrum of V339 Del obtained at +38 d.}
 Panel {\bf a} shows the radial velocity plots of three Fe\,{\sc ii}
 emission lines belonging to the same multiplet
 number$^{\mbox{\scriptsize 30}}$ (42).
 In addition to the similarity of their broad emission profiles,
 all lines have common blue-shifted absorption line features around
 $v_{\rm rad} \sim -1,000$ \kms.
 Panel {\bf b} shows the enlarged view of the absorption line
 features in panel {\bf a}.
 Dips of individual absorption line are indicated with dashed lines.
 {\bf c}, The absorption line systems in H\,{\sc i} Balmer lines drawn on the same velocity
 scale as in panel {\bf b}.
\\

\noindent
Extended Data Fig. 3:
 {\bf Near UV spectrum of V339 Del obtained at +38 d.}
 Panel {\bf a} shows the overall view of the spectrum from 308 to 350
 nm.
 Identified Fe\,{\sc ii} emission lines are indicated with red
 ticks at the bottom.
 The identified absorption line systems
 originating from iron-group ions 
 -- Fe\,{\sc ii} (red), Ti\,{\sc ii} (blue), Cr\,{\sc ii} (green),
 Mn\,{\sc ii}, Ni\,{\sc ii}, and V\,{\sc ii} (black) -- are indicated by
 ticks at the top.
 Panel {\bf b} shows a sample of the absorption line identification.
 The results of our identification are displayed along the spectrum.
 {\bf c}, Same as Extended Data Fig.\,2-{\bf b}, but for two lines
 (Ti\,{\sc ii} and Cr\,{\sc ii}) highlighted in panel {\bf b} are plotted
 on the velocity scale.
\\

\noindent
Extended Data Fig. 4:
 {\bf Spectra in the vicinity of the Be\,{\sc ii} doublet from +38 to +52 d.} 
 {\bf a} -- {\bf c},
 The horizontal scale is displayed with the heliocentric (bottom) and 
 the Doppler corrected  wavelengths (top).
 The Doppler corrections are applied using $v_{\rm days}$ 
 $= v_{+38}$, $v_{+47}$, and $v_{+48}$ for panels {\bf a}, {\bf b}, and
 {\bf c}, respectively.
 The local continuums fitted with high order (10--20) spline functions
 are over-plotted with green lines.
 The positions of the strongest 
 ($v_{\rm rad} =  v_{\rm days}$) and the second strongest components of
 the absorption line system are indicated by colored
 long and short lines connected by horizontal bars.
 {\bf d}, Since no apparent absorption lines are found in
 +52 d, the spectrum is  applied a Doppler correction using $v_{+48}$.

\begin{center}
\newpage
\includegraphics[width=17cm]{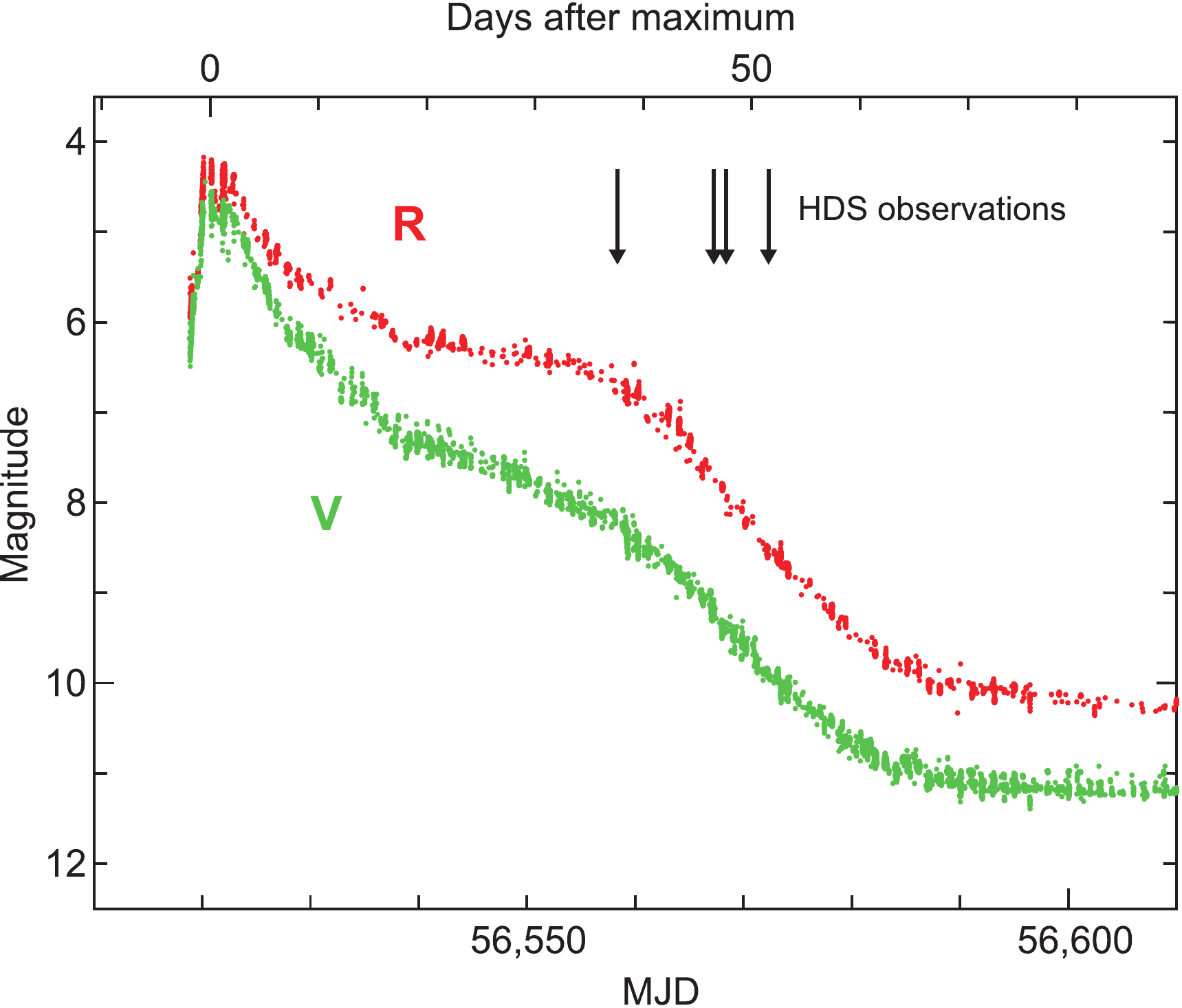}
\\
Extended Data Fig. 1:
 {\bf Optical light curves of V339 Del.}

\newpage
\includegraphics[width=17cm]{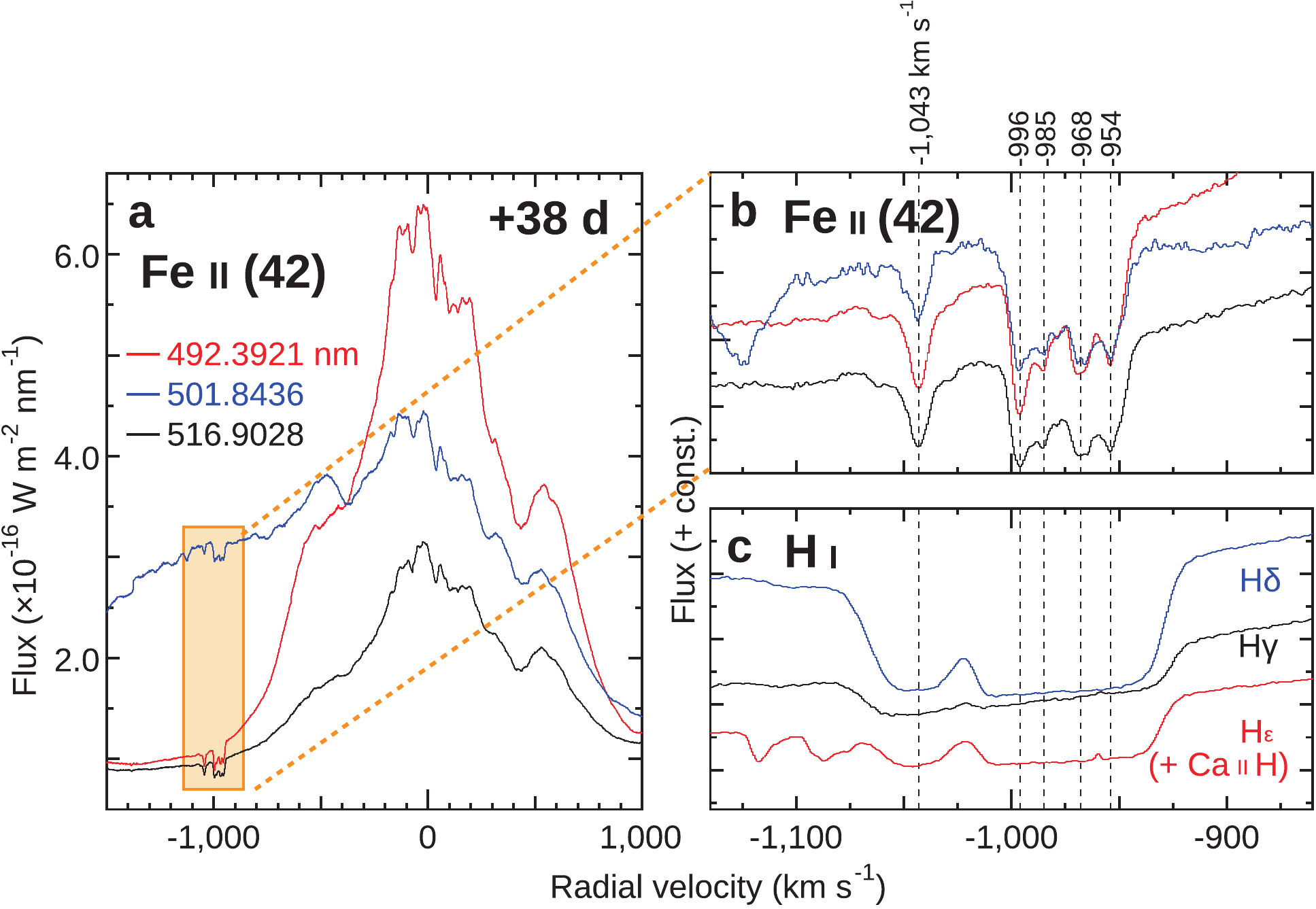}
\\
Extended Data Fig. 2:
 {\bf Optical spectrum of V339 Del obtained at +38 d.}

\newpage
\includegraphics[width=12.5cm]{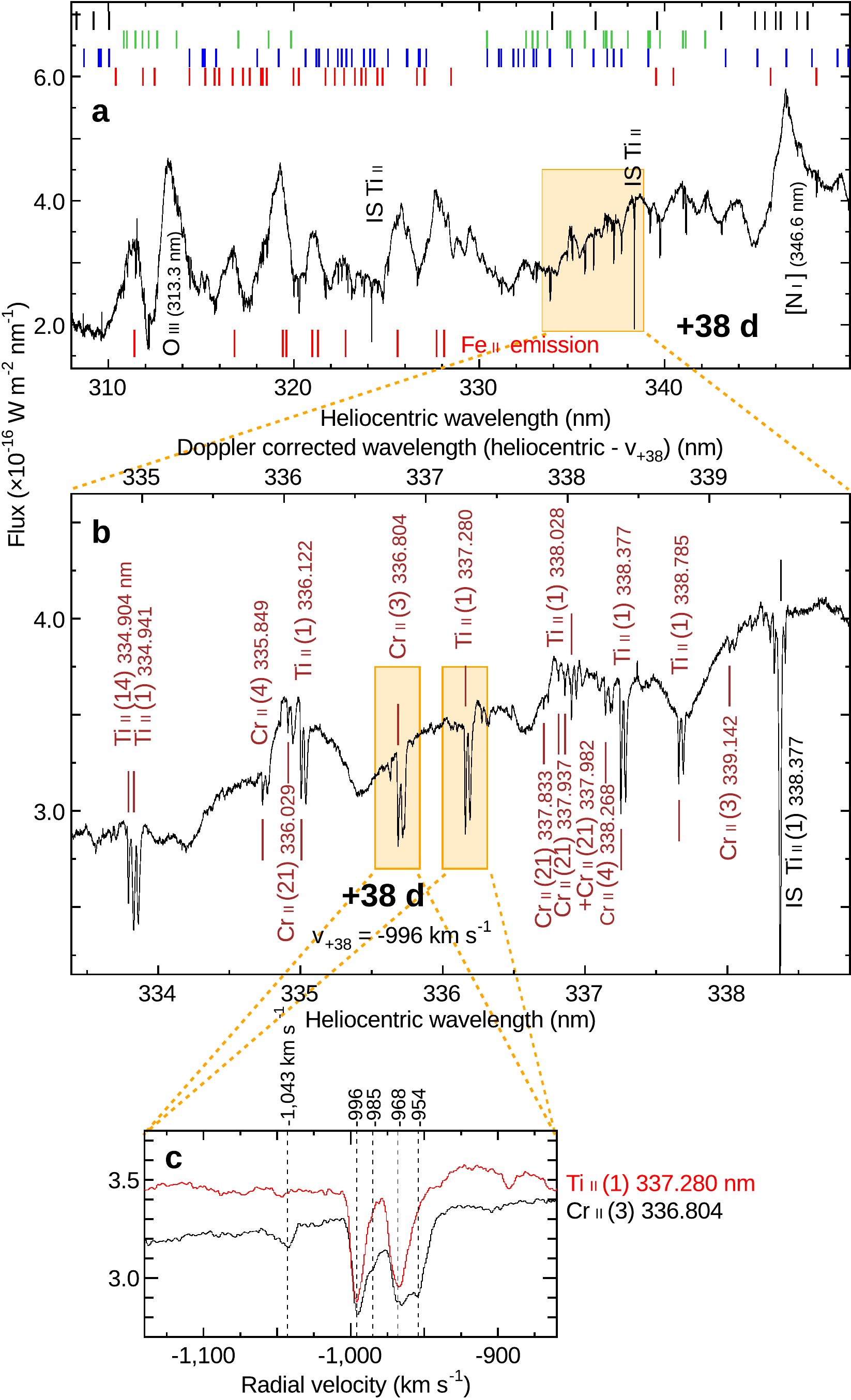}
\\
Extended Data Fig. 3:
 {\bf Near UV spectrum of V339 Del obtained at +38 d.}

\newpage
\includegraphics[width=13cm]{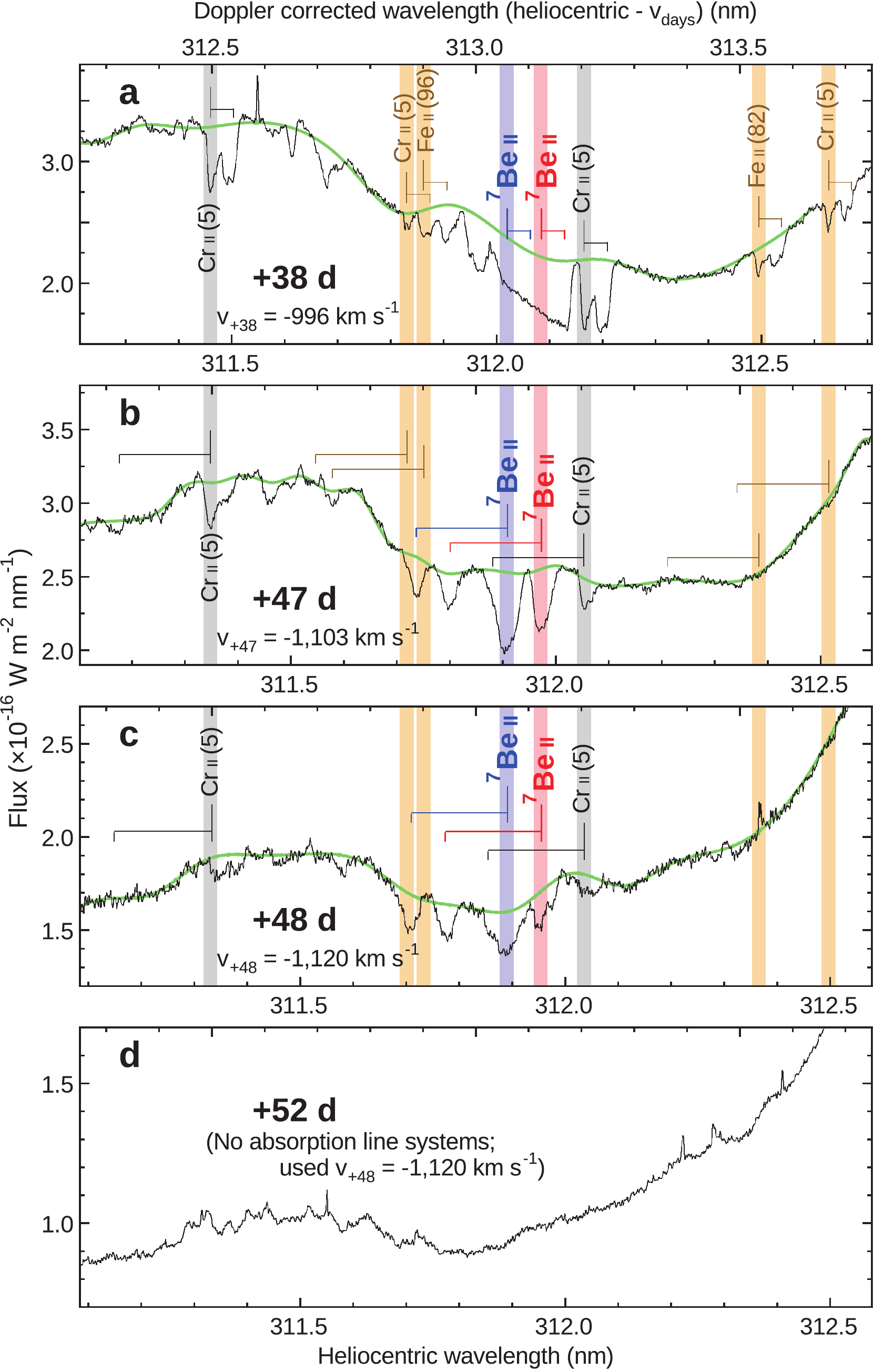}
\\
Extended Data Fig. 4:
 {\bf Spectra in the vicinity of the Be\,{\sc ii} doublet from +38 to +52 d.} 

\newpage
\noindent
Extended Data Table 1:
{\bf Journal of HDS observations of V339 Del}\\
\includegraphics[width=13cm]{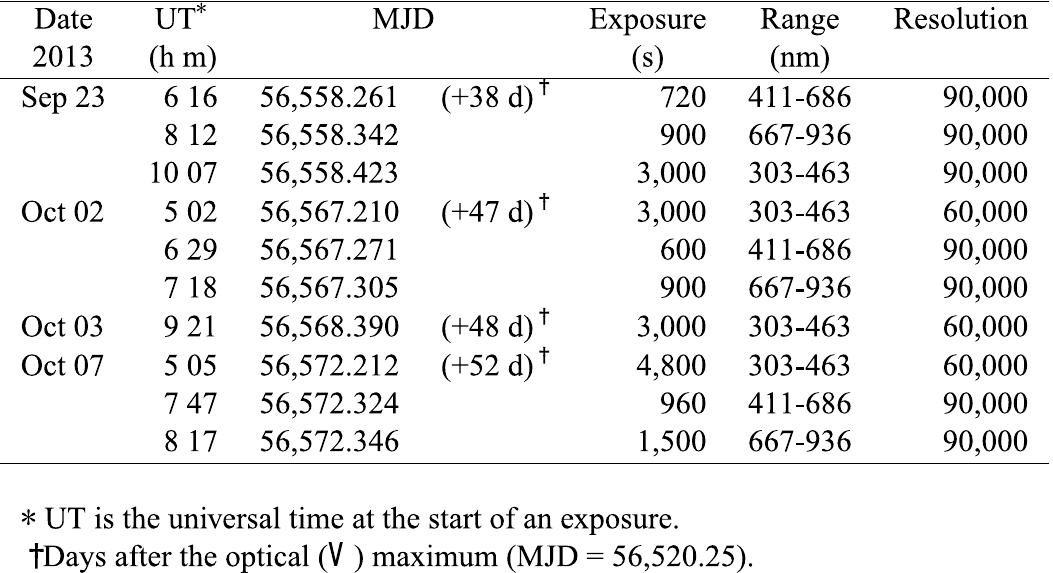}

\end{center}

\end{document}